\documentclass[iop]{emulateapj}

\usepackage{longtable}

\shorttitle{Methyl Acetate in Orion}
\shortauthors{Tercero et al.}

\begin{document}


\title{Discovery of Methyl Acetate and Gauche
Ethyl Formate in Orion
\thanks{This work was based on observations carried out with the 
IRAM 30-meter telescope. IRAM is supported by INSU/CNRS (France), 
MPG (Germany) and IGN (Spain).}
}

\author{B. Tercero}
\affil{Department of Astrophysics, CAB. INTA-CSIC. Crta Torrej\'on-Ajalvir, km. 4. 28850 Torrej\'on
de Ardoz. Madrid. Spain}
\email{terceromb@cab.inta-csic.es}

\author{I. Kleiner}
\affil{Laboratoire Interuniversitaire des Syst\`{e}mes Atmosph\'{e}riques, CNRS/IPSL UMR7583 et Universit\'{e}s Paris Diderot et 
Paris Est, 61 av. G\'{e}n\'{e}ral de Gaulle, 94010, 
Cr\'{e}teil, France}
\email{isabelle.kleiner@lisa.u-pec.fr}

\author{J. Cernicharo}
\affil{Department of Astrophysics, CAB. INTA-CSIC. Crta Torrej\'on-Ajalvir, km. 4. 28850 Torrej\'on
de Ardoz. Madrid. Spain}
\email{jcernicharo@cab.inta-csic.es}

\author{H. V. L. Nguyen}
\affil{Laboratoire Interuniversitaire des Syst\`{e}mes Atmosph\'{e}riques, CNRS/IPSL UMR7583 et Universit\'{e}s Paris Diderot et 
Paris Est, 61 av. G\'{e}n\'{e}ral de Gaulle, 94010, 
Cr\'{e}teil, France}
\email{nguyen@pc.rwth-aachen.de}

\author{A. L\'opez}
\affil{Department of Astrophysics, CAB. INTA-CSIC. Crta Torrej\'on-Ajalvir, km. 4. 28850 Torrej\'on
de Ardoz. Madrid. Spain}
\email{lopezja@cab.inta-csic.es}

\and
\author{G. M. Mu\~noz Caro}
\affil{Department of Astrophysics, CAB. INTA-CSIC. Crta Torrej\'on-Ajalvir, km. 4. 28850 Torrej\'on
de Ardoz. Madrid. Spain}
\email{munozcg@cab.inta-csic.es}

\begin{abstract}
We report on the discovery of methyl acetate, CH$_3$COOCH$_3$, through the detection of
a large number of rotational lines from each one of the spin states of the molecule: 
AA species (A$_1$ or A$_2$), EA species (E$_1$), AE species (E$_2$), EE species (E$_3$ or 
E$_4$). We also report the detection, for the first time in space,
of the $gauche$ conformer of ethyl formate, CH$_3$CH$_2$OCOH, in the same source.
The $trans$ conformer is also detected for the first time outside the galactic center 
source SgrB2.
From the derived velocity of the emission of methyl acetate we conclude that it arises mainly
from the compact ridge region with a total column density of (4.2$\pm$0.5)$\times$10$^{15}$
cm$^{-2}$. The derived rotational temperature is 150 K.
The column density for each conformer of ethyl formate, $trans$ and $gauche$, is (4.5$\pm$1.0)$\times$10$^{14}$
cm$^{-2}$. Their abundance ratio indicates a kinetic temperature of 135 K for the emitting gas
and suggests that gas phase reactions could participate efficiently in the formation of both conformers
in addition to cold ice mantle reactions on the surface of dust grains.
\end{abstract}


\section{Introduction}
The line survey of Orion carried out by Tercero and collaborators using the IRAM 30-m telescope
presents a forest of lines arising from isotopologues and vibrationally excited
states of abundant species (see, e.g., \citealt{Tercero2010,
Tercero2011, Daly2013}). The problem of identifying these features was a real challenge as
initially we had more than 8000 unidentified lines. The analysis of the data has been done
molecule by molecule \citep{Tercero2012}. For each species we explored the literature
for spectroscopic information on the isotopologues and vibrationally excited states but
substantial laboratory work was missing for most of these species. 
In 2006 we started a close collaboration
with different spectroscopic laboratories that allowed us to identify nearly 4000 of these
unknown lines (often called $weeds$). The $^{13}$C and $^{15}$N isotopologues from ethyl cyanide,
CH$_3$CH$_2$CN, were measured by \cite{Demyk2007} and \cite{Margules2009}. Several vibrationally
excited states of the same species were characterized in the laboratory
by \cite{Daly2013}. The $^{13}$C, $^{18}$O,
and deuterated isotopologues of methyl formate were observed in the laboratory by
\cite{Carvajal2009}, \cite{Margules2010}, and \cite{Tercero2012b}. The vibrationally excited 
state $\nu_{12}$ of formamide was measured in the laboratory by \cite{Motiyenko2012}. Finally,
L\'opez et al. (in preparation) have characterized in the laboratory several vibrationally excited states
of vinyl cyanide, CH$_2$CHCN, which have been identified in Orion together with all its isotopologues $^{13}$C,
$^{15}$N, and D. All these isotopologues and vibrationally excited states were detected in space for the
first time thanks to these new laboratory data.
In addition, the study of the survey was divided in the analysis of
different families of molecules: CS bearing species \citep{Tercero2010},
silicon bearing molecules \citep{Tercero2011}, SO and
SO$_2$ \citep{Esplugues2013}. Work on other species such as HCN, HNC, and HCO$^+$, 
CH$_3$CN, HC$_3$N and HC$_5$N, and organic saturated O rich species, is in progress (Marcelino et al., Bell et al., 
Esplugues et al., L\'opez et al. in preparation).

Although many of the still 4000 remaining U lines could belong to rare isotopologues
of complex organic molecules, we think we are ready now to start the search for new
molecular species. The study of a cloud such as Orion could provide important clues on the formation of
complex organic molecules on the grain surfaces and/or in the gas phase. A systematic
line survey with most $weeds$ removed permits to address the problem of the abundances
of isomers and derivates of key species, such as methyl formate, ethyl cyanide, and others.
Moreover, it will constitute the best spectral template for future ALMA observations of hot cores.

In this letter we report the detection for the first time in space 
of methyl acetate, CH$_3$COOCH$_3$,
from the detection of many lines from the states AA, AE, EA, and EE (E$_3$, and E$_4$) 
of this molecule at 3, 2, and 1.3 mm. The $gauche$ conformer of ethyl formate (an isomer of
methyl acetate), for
which the $anti$ conformer was previously detected in SgrB2 by \citet{Belloche2009}, 
has also been detected in Orion. This detection of ethyl formate outside
the galactic center indicates that this species is also efficiently produced in hot cores.

\begin{figure}
\includegraphics[angle=0,scale=.9]{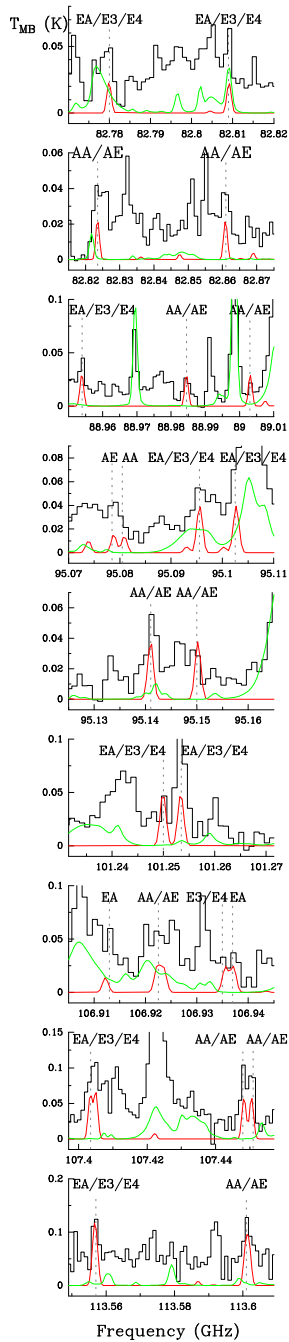}
\caption{Selected lines of Methyl Acetate at 3 mm, CH$_3$COOCH$_3$, 
towards Orion-IRc2.
The lines from the different states are identified. The continuous green line 
corresponds to all lines already modeled in our previous papers (see text).}
\end{figure}

\begin{figure}
\includegraphics[angle=0,scale=.9]{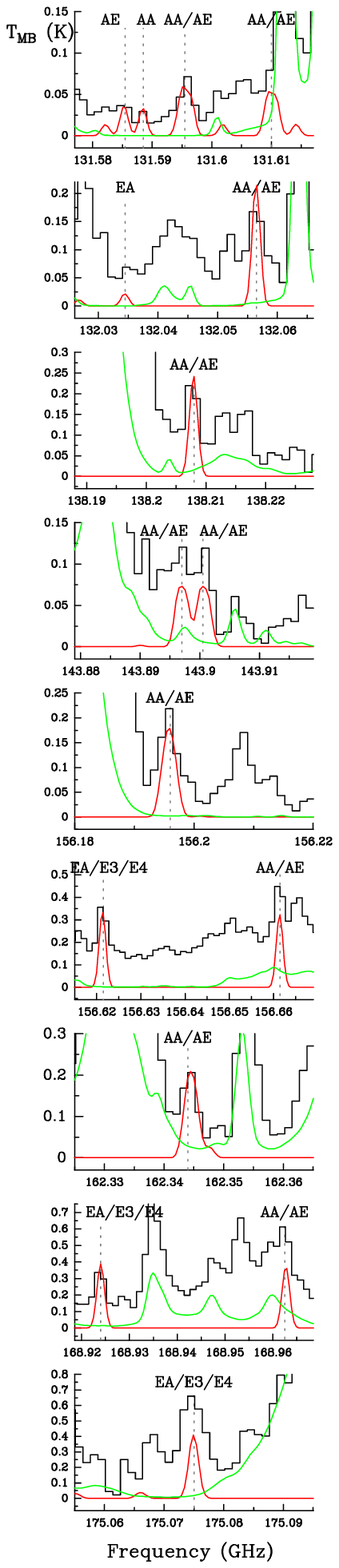}
\caption{Selected lines of Methyl Acetate at 2 mm, CH$_3$COOCH$_3$, towards Orion-IRc2.
The lines from the different states are identified. The continuous green line 
corresponds to all lines already modeled in our previous papers (see text).}
\end{figure}

\begin{figure}
\includegraphics[angle=0,scale=.9]{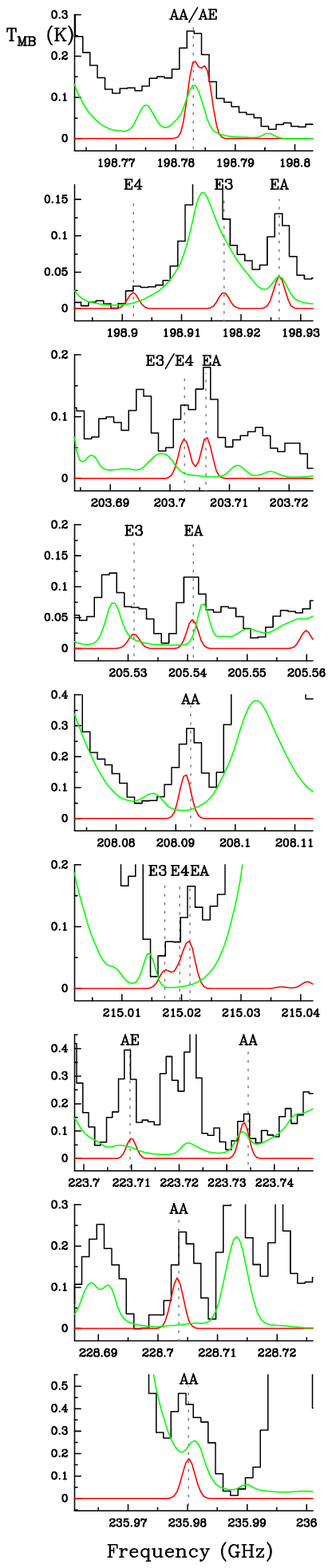}
\caption{Selected lines of Methyl Acetate at 1.3 mm, CH$_3$COOCH$_3$, towards Orion-IRc2.
The lines from the different states are identified. The continuous green line 
corresponds to all lines already modeled in our previous papers (see text).}
\end{figure}

\begin{figure}
\includegraphics[angle=0,scale=.85]{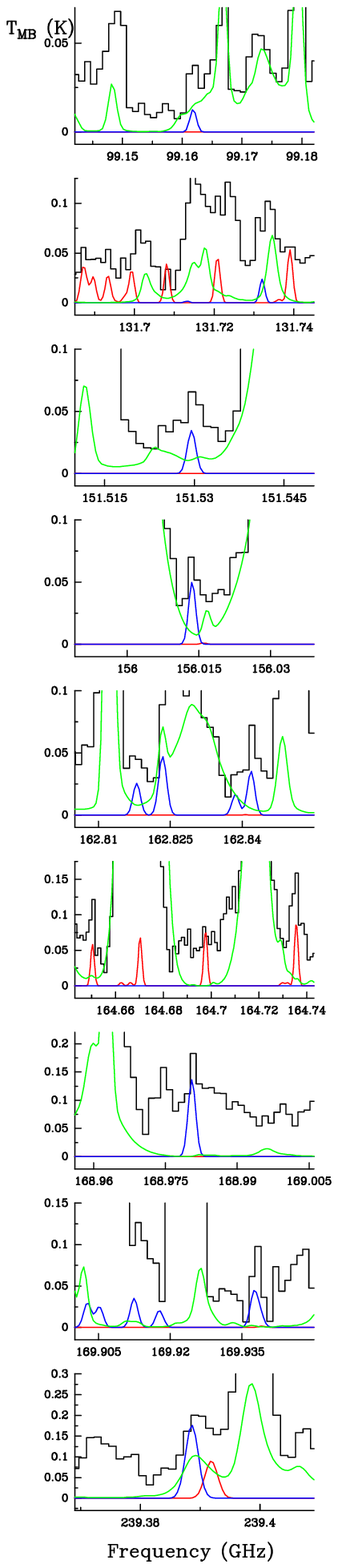}
\caption{Selected lines of the $trans$ (red line) and $gauche$ (blue) conformers
of ethyl formate, CH$_3$CH$_2$OCOH, towards Orion-IRc2. The synthetic spectrum corresponds
to the same column density for both conformers (see text).
The continuus green line 
corresponds to all lines already modeled in our previous papers (see text).}
\end{figure}

\section{Observations}
The observations were carried out using the IRAM 30-m radiotelescope during several 
periods (see \citealt{Tercero2010}). 
System temperatures were in the range 200-800 K for the 1.3 mm 
receivers, 200-500 K for the 2 mm receivers, and 100-350 K for the 3 mm receivers, depending 
on the particular frequency, weather conditions, and source elevation. 
The intensity scale was calibrated 
using two absorbers at different temperatures and using the Atmospheric Transmission Model 
(ATM, Cernicharo 1985; Pardo et al. 2001). 
Pointing and focus were regularly checked on the nearby quasars 0420-014 and 0528+134. 
Observations were made in the balanced wobbler-switching mode.
We pointed the observations towards IRc2 
at $\alpha$(J2000)=5$^{h}$ 35$^{m}$ 14.5$^{s}$, 
$\delta$(J2000)=-5$\degr$ 22$\arcmin$ 30.0$\arcsec$. 
The data were processed using the IRAM GILDAS software\footnote{http://www.iram.fr/IRAMFR/GILDAS} 
(developed by the Institute de Radioastronomie Millimetrique). We considered lines with 
intensities $\geq$0.02 K, covering three or more channels.
Figures are shown in units 
of main beam antenna temperature, T$_{MB}$=T$^{\star}_A$/$\eta_{MB}$,
where $\eta_{MB}$ is the main beam efficiency. A more detailed description
of the observations can be found in \cite{Tercero2012}.

\section{Frequency and Intensity Predictions for Methyl Acetate}

Methyl acetate [CH$_3$-O-C(=O)- CH$_3$], has been studied by \cite{Sheridan1980} using Stark 
spectrometers in the frequency region from 8 to 40 GHz. 
In 2011, the investigation 
was considerably improved with 315 new lines recorded using the molecular beam Fourier transform 
microwave spectrometer in Aachen and 519 lines recorded using the Free jet absorption 
Stark-modulated millimeterwave (FJASMmm) spectrometer in Bologna \citep{Tudorie2011}. A newly written program 
BELGI-Cs-2Tops based on the Hamiltonian described by \cite{Ohashi2004} was used to fit the complete 
data set with 27 molecular parameters, up to $J$ = 19 and $K_a$ = 7. More than 800 new microwave and 
millimeter-wave measurements were assigned to the ground-state transitions in methyl acetate and fit, leading to 
root-mean-square deviations of 4 kHz for the microwave lines and of 40 kHz for the millimeter-wave 
lines, i.e., to residuals essentially equal to the experimental measurement errors. The heights 
for two internal rotation barriers were determined to be of 102 cm$^{-1}$ for the acetyl CH$_3$ internal rotor and of 422 cm$^{-1}$  for the ester CH$_3$.
As the theoretical background and code presently used have been extensively described by \cite{Tudorie2011}, we 
will summarize here only the main characteristics. 

The BELGI-Cs-2Tops program is restricted to $i)$ asymmetric top molecules containing two non-equivalent 
CH$_3$ internal rotors, and $ii)$ molecules belonging to the Cs point group at equilibrium. 
This code is most closely related to a code used only once in the literature for a treatment
of the microwave spectrum of the molecule N-methylacetamide
[CH$_3$-NH-C(=O)-CH$_3$] \citep{Ohashi2004} although some differences exist as detailed in \cite{Tudorie2011}.

The torsional and rotational Hamiltonian is diagonalized in separate symmetry blocks, each characterized by one 
of the five ($\sigma_1$, $\sigma_2$) pairs, where $\sigma_1$ and $\sigma_2$ designates the symmetry indicators for each of the A and B tops. We 
follow the notation used in Table 1 of \cite{Ohashi2004}, i.e. we have for the 5 symmetry species: 
$\sigma_1$=0, $\sigma_2$=0, AA species (A$_1$ or A$_2$ in the G$_{18}$ permutation-inversion group of 
the molecule), $\sigma_1$=$\pm$1, $\sigma_2$=0, EA species (E$_1$), $\sigma_1$=0, $\sigma_2$=$\pm$1, 
AE species (E$_2$), $\sigma_1$=$\pm$1, $\sigma_2$=$\mp$1, EE species (E$_3$), 
$\sigma_1$=$\pm$1, $\sigma_2$=$\pm$1, EE species (E$_4$). The higher barrier 
hindering the ester methyl group internal rotation (422 cm$^{-1}$) corresponds to the smaller splittings between  
the AA and the E$_2$ lines, whereas the lower barrier (102 cm$^{-1}$) hindering the acetyl methyl group is 
responsible for the larger splittings between the AA and the E$_1$ lines. The splittings between the E$_3$ lines and the E$_4$ lines are due to the 
coupling between 
the two tops. Statistical weights for methyl acetate are 16, 16, 16, 8 and 8 for the A$_1$ or A$_2$ (AA), 
E$_1$ (AE), E$_2$ (EA), E$_3$, and E$_4$, respectively \citep{Ohashi2004}. 
The zero point energy for the $J$=$K$=0 levels are 99.9450 cm$^{-1}$, 
99.9551 cm$^{-1}$, 101.0947 cm$^{-1}$, 101.1047 cm$^{-1}$, and 101.1049 cm$^{-1}$ for A, E$_2$, E$_1$, E$_3$, and 
E$_4$ species, respectively. 

The spectroscopic constants determined from the previous fit published in \cite{Tudorie2011} are 
used in the present study to predict the transition frequencies for the $\nu_t$=0 torsional ground  
state up to $J$ = 30. Since we did only fit the observed lines in the laboratory up to $J$ = 19, $K_a\le$7, 
large uncertainties exist for the higher $J$ and $K_a$ values (we estimate 1$\sigma$ uncertainties of 0.1-0.5
MHz for $J<$25 and 0.5-2 for 25$\le$$J$$\le$30). For the present paper we 
also have implemented in the BELGI-Cs-2Tops the calculation of the intensities using the same method as 
described in \cite{Hougen1994,Kleiner2010} for the one-top codes.
The BELGI-Cs-2Tops code is written in a quasi principal-axis-method PAM axis system
which can be obtained by a rotation about the c-axis
from the principal axis system (PAM) by the angle $\theta$, as
shown in Eq. (9) of \cite{Ohashi2004}.  
The value of $\theta$ in methyl acetate is determined to be  -0.0316385(95) radians \citep{Tudorie2011}. 
Using the values determined by \cite{Sheridan1980} in the PAM axis system, we obtain 
for the components of the electric dipole moments in the quasi-PAM axis system,  $\mu_a$ = -0.008 Debye and 
$\mu_b$= 1.641 Debye. 

\section{Results and Discussion}
The predicted frequencies and intensities, together with the energy of the levels, have been 
implemented in the MADEX code \citep{Cernicharo2012}. Each one of the five states, AA, AE (E$_2$), 
EA (E$_1$), EE (E$_3$ and E$_4$) has been considered as an independent molecular species for the 
calculation of line intensities. 
For the kinetic temperature of Orion levels above $J$ = 30 could be significantly 
populated and some lines could also contribute to the Orion spectrum. The predictions from our
spectroscopic code have, however, large uncertainties for larger $J$'s (see above).  
The synthetic spectrum of methyl acetate was calculated assuming LTE conditions (due to the lack
of collisional rates for this molecule)
for a kinetic temperature of 150$\pm$20 K, a LSR velocity of the cloud of 8 km\,s$^{-1}$, a line width of 3
km\,s$^{-1}$, and a column density for each state AA, EA, and AE of (10.4$\pm$1.0)$\times$10$^{14}$ cm$^{-2}$, and of
(5.2$\pm$1.0)$\times$10$^{14}$ cm$^{-2}$ for the E$_1$ and E$_2$ states. Size (15'') and offset (7'') from the pointing position (IRc2) of the compact ridge component (see \citealt{Favre2011}, \citealt{Genzel1989}, \citealt{Blake1987}) are taken into account in our model.
Beam dilution is corrected for each line depending on their frequency.
Consequently, the total column density of methyl acetate in Orion is 4.2$\pm$0.5$\times$10$^{15}$ cm$^{-2}$ which
includes a correction factor to the partition
function computed for $T$$_K$=150 K and $J$$\le$30 of 2.6. It includes 
the torsional excited states (estimated energies) and rotational levels up to $J$=65. 
Figures 1, 2, and 3 show selected lines from the best fit synthetic spectrum to the Orion data. The full list of
detected lines is provided in Table 1 (electronic  version) where we list 215 unblended 
lines as well as 163 lines moderately blended with other species. No unblended lines of CH$_3$COOCH$_3$ are missing.
The detection is fully secure taking into account that the systematic pattern of the lines arising from the 
different
states is always present. Lines from $\nu_t$=1 could be also detectable as the vibrational partition function indicates
that the population of the $\nu_t$=1 levels from the two methyl groups will be a factor 1.5-2
below that of the $\nu_t$=0 states. Unfortunately, no accurate predictions are available for the torsionally
excited states. 

Ethyl formate has two conformers, $trans$ (also called $anti$) and $gauche$ with the latter being
65$\pm$21 cm$^{-1}$ (94 K) above in energy \citep{Riveros1967}. The $gauche$ conformer has two possible
orientations of the terminal methyl group and, hence, it could be twice as abundant as the
$trans$ conformer if the energy difference were zero \citep{Belloche2009}. 
We have used the spectroscopic laboratory data from \cite{Riveros1967}, \cite{Meyer1970}, and \cite{Demaison1984} to derive
the rotational
constants which were incorporated into the MADEX code. The predictions were checked against the corresponding
entry in the JPL catalog \citep{Pickett1998}.
The $trans$ conformer of ethyl
formate was detected in SgrB2 by \cite{Belloche2009} but only upper limits were obtained for the
$gauche$ one. We have identified both conformers in the line survey of Orion trhough 90 features free of blending (52 $trans$ and $38$ $gauche$).
Selected identified lines are shown in Fig. 4.
We found that the $gauche$ conformer has less intense lines than the $trans$ one for
transitions at 3 and 2 mm. However, at higher frequencies there are many multiplet
transitions of the $gauche$ that coincide in frequency and they become prominent in the Orion data. The
best fit to the data (assuming the same physical conditions and source parameters 
than those of methyl acetate corresponding to the compact ridge component of Orion KL)
provide a rotational temperature of 150$\pm$20 K, and a 
column density for each conformer of (4.5$\pm$1.0)$\times$10$^{14}$ cm$^{-2}$, i.e., a total
column density for ethyl formate $\simeq$5 times below that of methyl acetate. 
From the observed abundance ratio between the
$trans$ and $gauche$ conformers, N$_{gauche}$/N$_{trans}\simeq$1= 2$\times$e$^{-94/T}$, 
we derive a kinetic temperature for the emitting gas of 135$\pm$30 K.
If the lowest energy conformer of ethyl formate, the $trans$ one, was formed
on the ices before the warm phase of the cloud, isomerization to the $gauche$ form
will require a high kinetic energy and, perhaps, a long time to overpass the barrier to isomerization
of 550 K \citep{Riveros1967}. Our result points towards a fast equilibrium between both
conformers at 150 K and in a time comparable to the duration of the present warm phase
of the cloud. If the molecule is formed in gas phase the energy liberated in the process
could help in the isomerization process and the observed conformer temperature will reflect
the kinetic temperature of the gas. However, no chemical paths are included in the present
chemical models to form neither ethyl formate nor methyl acetate. New laboratory experiments
are needed to understand the way these species could be formed in gas phase or in ices
and how conformers of the same species can equilibrate at a temperature close to the
kinetic temperature of the gas.

Methyl acetate is probably formed via multiple reaction pathways from species
detected in hot cores such as methyl formate, acetic 
acid, and methanol. In icy grain mantles, methanol is one of the common molecular components. 
Acetic acid is a relatively large molecule that would 
be difficult to detect in the ice by IR observations and that could be the precursor of methyl 
acetate and ethyl formate. 
Only formic acid (HCOOH) has been proposed as a possible carrier of the 7.24 $\mu$m band 
toward high-mass protostars, while the 7.41 $\mu$m band could be due to the formate ion (HCOO$^-$) 
and acetaldehyde (CH$_3$CHO), 
according to \cite{Schutte1999}.  Formation of HCOOH on a surface occurs experimentally at low temperatures, 
mainly through hydrogenation of the HO-CO complex \citep{Ioppolo2011}. Other possible formation routes in 
the ice are via precursor cations \citep{Woon2011} or by reactions of superthermal O($^3P$) atoms and CH$_4$ with an over coat of 
CO \citep{Madzunkov2010}. Also photon or electron irradiation of H$_2$O:CO ice mixtures leads to formation 
of formic acid among other products, including methanol in the case of photoprocessing 
\citep{Watanabe2007,Bennett2011}. 
HCOOH was found to spontaneously deprotonate when sufficient water is present to stabilize charge transfer complexes. 
Both ammonia and water can serve as proton acceptors, yielding ammonium (NH$_4^+$) and hydronium (H$_3$O$^+$) counterions 
\citep{Park2006}. The so-formed formate ion (HCOO$^-$) might intervene in the formation of species 
like methyl acetate in the ice matrix, but this was to our knowledge not confirmed experimentally. \cite{Brouillet2013} have recently
observed with high angular resolution CH$_3$OCH$_3$ and CH$_3$OCOH towads Orion-IRc2 and conclude that the similarity in the spatial
distribution of both species points towards a common precursor. The observation of a similar abundance for the 
two
conformers of ethyl formate points to a gas phase production path rather than to a low temperature ice formation mechanism. Radicals
such as methoxy (CH$_3$O, \citealt{Cernicharo2012}) or CH$_3$CO could play an important role in the gas phase chemistry. However,
methoxy has been observed only towards cold dark clouds \citep{Cernicharo2012b} and it is not 
detected in our line survey of Orion; CH$_3$CO has not been detected yet in the ISM.

The two conformers of ethyl formate and 
methyl acetate are isomers of the C$_3$H$_6$O$_2$. Three additional isomers, propanoic (propionic)
acid (CH$_3$CH$_2$COOH),
hydroxyacetone (CH$_3$COCH$_2$OH), and methoxyacetaldehyde (CH$_3$OCH$_2$OH), could be also present in Orion. 
The three species are implemented in MADEX. For hydroxyacetone the available spectroscopic data 
have been summarized by \cite{Braakman2010} and cover frequencies up to 431.8 GHz (dipole moments from 
\citealt{Kattija-Ari1980}). We obtain an upper limit to its column density of 8$\times$10$^{13}$ cm$^{-2}$
(see also \citealt{Apponi2006} for an upper limit to its column density towards SgrB2).
For propionic acid and methoxyacetaldehyde we also obtain upper limits to their column densities of 
1.6$\times$10$^{14}$ and 2$\times$10$^{14}$ cm$^{-2}$, respectively. We note, however, 
that frequency predictions above 40 GHz for these two molecular
species are rather uncertain \citep{Stifvater1975,Ouyang2008,Hirano1987}. Hence, of the known 
possible non cyclic isomers
of C$_3$H$_6$O$_2$, methyl acetate appears to be the most abundant one. Laboratory spectroscopic data are needed
for propionic acid and methoxyacetaldehyde in order to draw further conclusions on their contribution to
the ice mantle and gas phase chemistry of hot cores, and to the forest of still unknown spectral features 
in Orion.

\scriptsize
\begin{table*}
\begin{center}
\caption{Detected lines of CH$_3$COOCH$_3$} \label{tab_ch3cooch3}
\resizebox{1\textwidth}{!}{%
\begin{tabular}{rrrrrrrrcrrrrrrl}
\hline\hline 
$J$ & $K_a$ & K$_c$ & $p$ & $J'$ & $K'_a$ & $K'_c$ & $p'$ & State & Predicted & Error  & $E_u$    & $S_{ij}$ $\mu^2$ & Observed  & $T_{mb}$  & Blend\\
 & & & & & & & & & freq. (MHz) & (MHz) & (K) & & freq. (MHz) & (K) & \\
\hline \hline
  13&   0&  13&    &  12&   1&  12&    & \tiny{E3} &  82779.635&   .012 &   28.6 & 31.30 & 82780.2    &0.03  & \tiny{CH$_3$CH$_2$CN}                                           \\
& & & & & & & & & & & & & & & \tiny{$^{33}$SO$_2$} \\ 
  13&   0&  13&    &  12&   1&  12&    & \tiny{E4} &  82779.725&   .012 &   28.6 & 31.30 & $\dagger$  &      &                                                                                         \\
  13&   0&  13&    &  12&   1&  12&    & \tiny{EA} &  82780.026&   .012 &   28.6 & 31.30 & $\dagger$  &      &                                                                                         \\
  13&   1&  13&    &  12&   0&  12&    & \tiny{E4} &  82808.862&   .012 &   28.6 & 31.30 & 82809.2    &0.04  & \tiny{HCOOCH$_3$}                                                           \\
& & & & & & & & & & & & & & &   \tiny{$\nu_t$=1} \\ 
  13&   1&  13&    &  12&   0&  12&    & \tiny{E3} &  82808.881&   .012 &   28.6 & 31.30 & $\dagger$  &      &                                                                                         \\
  13&   1&  13&    &  12&   0&  12&    & \tiny{EA} &  82809.267&   .012 &   28.6 & 31.30 & $\dagger$  &      &                                                                                         \\
  13&   0&  13&    &  12&   1&  12&    & \tiny{AE} &  82823.351&   .012 &   28.6 & 31.30 & 82823.2    &0.04  &                                                                                         \\
  13&   0&  13&   1&  12&   1&  12&   1& \tiny{AA} &  82823.702&   .012 &   28.6 & 31.30 & $\dagger$  &      &                                                                                          \\
\hline  
\end{tabular}
}
\end{center}
Continue in the electronic version.\\
{\footnotesize{\sf{
{\bf{Note.-}} Emission lines of CH$_3$COOCH$_3$  
present in the spectral scan of Orion KL from the   
IRAM 30-m radio-telescope. Column 1-8 indicate the line transition, Col. 9 the state of the molecule, Col. 10   
gives the predicted frequency in the laboratory, Col. 11 uncertainty of frequency predictions, Col. 12 upper level energy, Col. 13 line strength,
Col. 14 observed frequency assuming a v$_{LSR}$ of 8 km s$^{-1}$, Col. 15 mean beam temperature, and Col. 16 blends.\\  
$\dagger$ Blended with previous line. \\                                         
}}}
\end{table*}


\acknowledgments
J. Cernicharo, B. Tercero, G. Mu\~noz Caro, and A. L\'opez thank the Spanish MICINN for funding support through grants, AYA2009-
07304, AYA2011-29375, and CSD2009-00038. I. Kleiner acknowledges support from the french national PCMI program, and support 
from the ANR-08-BLAN-0054 contract (France).
J. Cernicharo thanks U. Paris Est for an invited professor position associated to this work.

\clearpage

\end{document}